\documentclass[% 
reprint,twocolumn,
 amsmath,amssymb,
 aps,
prl,
]{revtex4}
\usepackage{graphicx}
\usepackage{epstopdf}

\newcommand{\ve}[1]{\mathbf{#1}}
\newcommand{\lap}{\Delta}
\newcommand{\av}[1]{\left\langle#1\right\rangle}
\newcommand{\edd}{\epsilon_{\text{dd}}}
\newcommand{\nhm}{n_{\text{HM}}}

\begin{document}

%\preprint{APS/123-QED}
\title{Dipolar Bose-Einstein Condensates with Weak Disorder}
\author{Christian Krumnow}
\affiliation{Institut f\"{u}r Theoretische Physik, Freie Universit\"{a}t Berlin,
Arnimallee 14, DE-14195 Berlin, Germany}
\author{Axel Pelster}
\affiliation{Fakult\"at f\"ur Physik, Universit\"at Bielefeld,
Universit\"atsstra{\ss}e 25,
DE-33501 Bielefeld, Germany, and\\ Fachbereich Physik, Universit\"{a}t
Duisburg-Essen, Lotharstra{\ss}e 1, DE-47048 Duisburg, Germany}
\date{\today}
\begin{abstract}
A homogeneous polarized dipolar Bose-Einstein condensate is considered in the
presence of weak quenched disorder within mean-field theory at zero temperature. By
first solving perturbatively the underlying Gross-Pitaevskii equation and then
performing disorder ensemble averages for physical observables, it is shown that the
anisotropy of the two-particle interaction is passed on to both the superfluid
density and the sound velocity.
\end{abstract}
\maketitle
Bose-Einstein condensates (BECs) in a disordered environment have been the subject
of various experimental investigations
in recent years. Historically, this ``dirty boson'' problem arose in the context of
superfluid helium in Vycor 
glass \cite{reppy}. Later, disorder either appeared naturally, for instance in
magnetic wire traps \cite{krueger,fortagh}, 
or was created artificially and controllably by using laser speckles
\cite{dainty,aspect}.
Most theoretical studies focused on analyzing the Bogoliubov theory of dirty bosons
for weak disorder within
either the second quantization \cite{huang,giorgini,kobayashi,falco1} or the replica
method \cite{vinokur,dresden}.
In this way, it turned out that superfluidity persists despite quenched randomness,
but a depletion occurs due to
the localization of tiny condensates in the respective minima of the disorder
potential. So far, this
localization scenario has only been analyzed for the contact interaction, as it is
usually dominant for ultracold
dilute Bose gases. Since the realization of atomic dipolar BECs \cite{pfau}
and the generation of heteronuclear molecules in the rovibrational ground state near
quantum degeneracy
\cite{ospelkaus},
long-range and anisotropic dipole-dipole interactions have also attracted attention
\cite{baranov,carr,santos}.
Therefore, we consider in this Rapid Communication the impact of weak disorder upon
a polarized dipolar BEC
at zero temperature. In particular, we will show that both the superfluid
density and the sound velocity yield characteristic interaction-induced
anisotropies, which are not present at zero temperature in the absence of disorder.

The weakly interacting theory of dirty bosons
\cite{huang,giorgini,kobayashi,falco1,vinokur,dresden} states 
that Bogoliubov quasiparticles and disorder-induced fluctuations decouple in the
lowest order. This suggests 
a simplified approach in which the leading correction due to the presence of a
random potential could be derivable
from a mean-field theory. Therefore, we assume at $T=0$ that all bosons occupy the
same quantum state,
for which the macroscopic wave function $\Psi(\ve{x})$ obeys
the time-independent Gross-Pitaevskii (GP) equation
\begin{eqnarray}
\left[-\frac{\hbar^2}{2m}\lap +U(\ve{x})+\int
d^3x^\prime\left|\Psi(\ve{x}^\prime)\right|^2V_{\text{int}}(\ve{x}
-\ve{x}^\prime)\right]\Psi(\ve{x})\nonumber \\ 
= \mu  \Psi(\ve{x})\, , \hspace*{1cm}
\label{GP}
\end{eqnarray}
and neglect with this from now on any impact of quantum fluctuations.
Here, $m$ denotes the particle mass, $\mu$ stands for the chemical potential, 
$V_{\text{int}}(\ve{x} -\ve{x}^\prime)$ represents an arbitrary two-particle
potential with inversion symmetry,
and $U(\ve{x})$ describes the disorder potential, which is defined by its
statistical properties. Denoting the disorder 
ensemble average according to $\av{\,\bullet\,}$, a homogeneous disordered system
has a vanishing 
first moment $\av{U(\ve{x})} = 0$ and the second moment is of the form 
$\av{U(\ve{x})U(\ve{x}^\prime)} = R(\ve{x}-\ve{x}^\prime)$.
The GP equation (\ref{GP}) represents a stochastic nonlinear partial differential
equation, where the
given statistics of the disorder potential $U(\ve{x})$ are mapped to the wave
function $\Psi(\ve{x})$ \cite{navez}.
In the case that the random potential $U(\ve{x})$ is small in comparison with all
other energy scales, its perturbative treatment is justified.
To this end, we decompose the wave function of the system according to
\begin{eqnarray}
\label{expansion}
\Psi(\ve{x}) = \psi_0 (\ve{x}) + \psi_1(\ve{x}) + \psi_2(\ve{x})+\ldots\,,
\end{eqnarray}
and solve the GP equation in the zeroth, first, and second order of $U(\ve{x})$,
respectively. 
As the expansion (\ref{expansion}) determines the ground state, the wave function
$\Psi(\ve{x})$ turns out to be real.
Afterwards, we determine the disorder ensemble average for both the particle density 
$n = \av{\Psi(\ve{x})^2}$ and the condensate density $n_0 = \av{\Psi(\ve{x})}^2$,  
thus, the condensate depletion  results in the lowest order in 
\begin{equation}
n-n_0 = n\int \frac{d^3k}{(2\pi)^3}\frac{R(\ve{k})}{\left[ \hbar^2 \ve{k}^2 / 2m
+2n V_{\text{int}}(\ve{k})\right]^2} + \ldots \,.
\label{conddep}
\end{equation}
Note that the existence of the ${\bf k}$ integral implies that the Fourier transform
of the interaction potential
$V_{\text{int}}(\ve{k})$ has to be strictly positive.
Physically, this condensate depletion is due to the formation of fragmented
condensates in the respective minima
of the random potential. In order to further 
quantify this notion, a separate order parameter was recently proposed in
Ref.~\cite{graham}. It is motivated
by defining the condensate density $n_0$ as usual as the superfluid order parameter
from the
off-diagonal long-range order of the one-particle density matrix \cite{leggett}
\begin{eqnarray}
\label{n0}
\lim_{|\ve{x}-\ve{x^\prime}| \rightarrow \infty} \av{\Psi(\ve{x}) \Psi
(\ve{x^\prime})} = n_0 \, .
\end{eqnarray}
The density of fragmented condensates $q$ is then identified as a separate
Bose-glass order parameter,
similar to the Edwards-Anderson order parameter of a spin glass \cite{edwards}. It
can be defined by considering the 
off-diagonal long-range order of the two-particle density matrix
\begin{eqnarray}
\label{q}
\lim_{|\ve{x}-\ve{x^\prime}|\rightarrow \infty} \av{\Psi(\ve{x})^2 \Psi
(\ve{x^\prime}) ^2} = (n_0+q)^2\,,
\end{eqnarray}
as the latter has to coincide with the square of the total density $n$.
Applying the concept of Eqs.~(\ref{n0}) and (\ref{q}) 
to the perturbative solution (\ref{expansion}) of the GP equation (\ref{GP}) indeed
yields,
together with Eq.~(\ref{conddep}), the result that the density of the fragmented
condensates $q$ defined in Eq.~(\ref{q}) coincides in the lowest
order with the condensate depletion $n - n_0$ found in Eq.~(\ref{conddep}). Thus, we
conclude that the localization phenomenon for weak
quenched disorder follows already from a mean-field description of the dirty boson
problem. 
Therefore, our mean-field approach 
represents a simplified derivation
for the disorder-induced condensate depletion (\ref{conddep}) 
in comparison with the Bogoliubov theory of
Refs.~\cite{huang,giorgini,kobayashi,falco1,vinokur,dresden}.
Note that, recently, disorder effects for Bogoliubov quasiparticles
have been analyzed in Ref.~\cite{gaul1}.

For a polarized dipolar BEC, the particles interact via a contact interaction with
strength 
$g = 4\pi\hbar^2 a / m$, where $a$ denotes the s-wave scattering length,
and they possess dipole moments, which are aligned along the $z$ axis. Thus,  the
Fourier
transform of the interaction potential is given by
$V_{\text{int}}(\ve{k}) = g[1 + \epsilon_{\text{dd}}(3x^2-1)]$,
where we define $x = \hat{\ve{k}}\cdot\hat{\ve{e}}_z= \cos \vartheta$ and 
$\edd= C_\text{dd} / 3g$ with $C_\text{dd}=\mu_0\ve{m}^2$, in the case of magnetic
dipoles with the dipole moment 
$\ve{m}$, and $C_\text{dd}=4\pi\ve{d}^2$ for electric dipoles, where the dipole
moment $\ve{d}$ is measured in Debyes. In the case of a spatially decaying disorder
correlation $R(\ve{x})$, the results of the theory do not depend significantly
on its shape \cite{timmer}. Therefore, in what follows, we restrict ourselves to the
case of a Gaussian
correlation with the Fourier transform 
$R(\ve{k}) = R \,e^{-\sigma^2\ve{k}^2 / 2}$, 
where
$R$ and $\sigma$ characterize
the strength and the correlation length of the disorder, respectively.
With this, the condensate depletion (\ref{conddep}) specializes to
$n-n_0  =  \nhm\ f\left(\edd,\sigma / \xi \right) + \ldots$,
where the limit of pure contact interaction and delta-correlated disorder yields 
$n_\text{HM}= [m^2 R / (8 \pi^{3/2} \hbar^4)] \,\sqrt{n / a}$ \cite{huang}.
Whereas the relative interaction strength $\edd$ increases the condensate depletion,
the
ratio of the correlation length $\sigma$ and the coherence 
length $\xi = 1/ \sqrt{8 \pi n a}$ decreases it according to 
the function 
\begin{eqnarray}
f\left(\edd,\frac{\sigma}{\xi}\right) = \int\limits_{0}^1dx \,
\frac{[ 1 +2\zeta (x)] e^{\zeta (x)}\text{erfc}\sqrt{\zeta(x)} }{\sqrt{3\edd
x^2+1-\edd} } -\frac{2\sigma}{\sqrt{\pi}\xi},\notag\\
\label{gisodip}
\end{eqnarray}
\begin{figure}[t]
\centering
\includegraphics[scale=0.65]{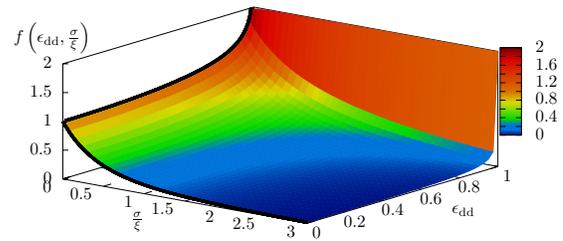}
\caption{(Color online) Dipolar condensate depletion function (\ref{gisodip}) with 
the special cases 
of vanishing correlation length $f(\edd,0)$ 
and pure contact interaction $f(0,\sigma / \xi)$ \cite{kobayashi}
indicated by solid lines.}
\label{fig:isoedd}
\end{figure}
\hspace*{-1.5mm}with the abbreviation $\zeta(x)= \sigma^2(3\edd x^2+1-\edd)/ \xi^2$
(see Fig.~\ref{fig:isoedd}).
In particular,
for small $\edd$, we have $f\left(\edd, \sigma / \xi \right) = A (\sigma / \xi) + B
(\sigma / \xi)\,\edd^2 + \ldots$,
with positive coefficients $A (\sigma / \xi)$ and $B (\sigma / \xi)$ which decrease
with increasing disorder correlation length.
In contrast, the condensate depletion function $f\left(\edd, \sigma / \xi \right)$
diverges in the limit
$\edd \uparrow 1$, in accordance with the conclusion that the existence of the ${\bf
k}$ integral in Eq.~(\ref{conddep}) is
lost provided $V_{\text{int}}(\ve{k})$ vanishes for any ${\bf k}$. 
This indicates that our perturbative treatment breaks down provided that the dipolar
interaction
is strong enough relative to the contact interaction.

\begin{figure*}[ht]
\centering
\includegraphics[scale=0.95]{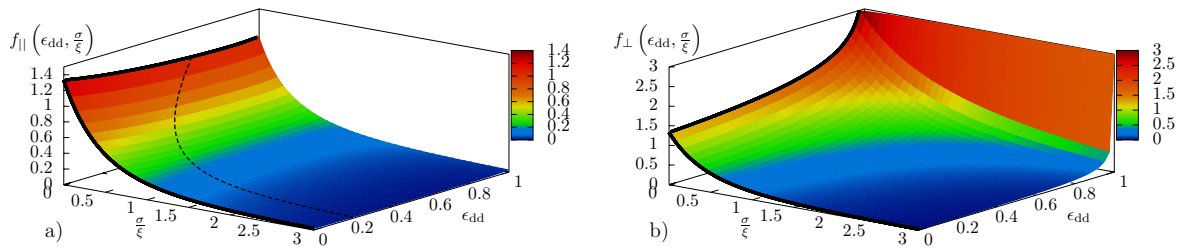}
\caption{(Color online) Superfluid depletion function a) parallel and b)
perpendicular to the polarization axis of the dipoles. The
solid lines indicate the special cases 
of vanishing correlation length  
and pure contact interaction, respectively. The dashed line defines the critical
relative interaction
strength via $f\left(\epsilon_{\text{dd,crit}}, \sigma / \xi \right)
=f_{||}\left(\epsilon_{\text{dd,crit}}, \sigma / \xi \right)$.}
\label{sup}
\end{figure*}
In order to describe a superfluid within mean-field theory, we apply a Galilean
boost with wave vector 
$\ve{k}_\text{N}$ to the time-dependent GP equation and introduce a moving
condensate via the 
ansatz $\Psi(\ve{x}) = \psi(\ve{x})e^{i\ve{k}_\text{S}\cdot\ve{x}}$ \cite{ueda}.
With this, we obtain in the stationary case
\begin{eqnarray}
&& \hspace*{-1cm}\Biggl[-\frac{\hbar^2}{2m}\lap
-i\frac{\hbar^2}{m}\ve{K}\cdot\nabla+U(\ve{x})-\mu_{\text{eff}}\nonumber\\
&& +\int d^3x^\prime \,V_{\text{int}}(\ve{x}-\ve{x}^\prime) |\psi(\ve{x}^\prime)|^2
\Biggr]\psi(\ve{x})
=0,\label{GPboostrw}
\end{eqnarray}
where we have introduced the abbreviations
$\ve{K} = \ve{k}_\text{S}-\ve{k}_\text{N}$ and
$\mu_{\text{eff}} =  \mu-\hbar^2 \ve{k}_\text{S}^2 / 2m +\hbar^2
\ve{k}_\text{N}\cdot\ve{k}_\text{S}/ m$.
Within a linear response, we can assume $\ve{k}_\text{N}$ and $\ve{k}_\text{S}$ to
be small,
so that the superfluid part of the system 
will move with a wave vector proportional to $\ve{k}_\text{S}$ and the nonsuperfluid
part is coupled 
to $\ve{k}_\text{N}$. Therefore, the expansion of the total momentum 
$\ve{P} = -\int d^3x\Psi^\ast(\ve{x})i\hbar\nabla\Psi(\ve{x})$ with respect to
$\ve{k}_\text{S}$ and 
$\ve{k}_\text{N}$ yields in the disorder-averaged case
$\av{\ve{P}} = V \left(
n_\text{S}\hbar\ve{k}_\text{S}+n_\text{N}\hbar\ve{k}_\text{N} \right)+\ldots\,$,
where $V$ 
denotes the volume of the system and $n_\text{S}$ and $n_\text{N}$ represent the
superfluid and the nonsuperfluid 
density, respectively. 
Expanding the solution $\psi(\ve{x})$ of Eq.~(\ref{GPboostrw}), which may now be
complex, 
with respect to the disorder
yields in general the result that the superfluid density turns out to be a tensor
with the components
\begin{eqnarray}
n_{\text{S},ij}  =  n \delta_{ij}
 -\hspace*{-1mm}\int\hspace*{-1mm}\frac{d^3k}{(2\pi)^3}\frac{4 nR(\ve{k})
k_i k_j}{\ve{k}^2\left[ \hbar^2\ve{k}^2 / 2m +2nV_{\text{int}}(\ve{k})\right]^2}
+\ldots 
\label{ns}
\end{eqnarray}
and $n_\text{S}+n_\text{N}=n$.
In the case of isotropy, i.e. $R(\ve{k}) = R(|\ve{k}|)$ and $V_\text{int}(\ve{k}) =
V_\text{int}(|\ve{k}|)$,
this reduces to  a diagonal superfluid density
$n_\text{S} = n-4 (n-n_0)/3+\ldots$, which generalizes the case of pure contact
interaction 
\cite{huang,giorgini,kobayashi,falco1,vinokur,dresden}.

For a dipolar BEC and Gaussian disorder correlation function, Eq.~(\ref{ns}) yields
a superfluid density 
that depends on the direction of the superfluid motion with respect to the
orientation of the dipoles.
Parallel to $\hat{\ve{e}}_z$, the superfluid depletion reads
$n - n_{\text{S},||} = \nhm f_{||}\left(\edd, \sigma / \xi \right) + \ldots$ with
the function
\begin{equation}
f_{||}\left(\edd,\frac{\sigma}{\xi}\right) =  4\int\limits_{0}^1dx 
\frac{x^2[1+2\zeta (x)]
e^{\zeta (x)}\text{erfc}\sqrt{\zeta (x)}}{\sqrt{3\edd x^2+1-\edd}}-\frac{8
\sigma}{3\sqrt{\pi}\xi},
\label{gisosuppar}
\end{equation}
whereas, perpendicular to the dipoles, we have $n - n_{\text{S},\bot} 
= \nhm f_{\bot} \left(\edd, \sigma / \xi \right)+\ldots$ with the function
$f_\bot\left(\edd, \sigma / \xi \right) =  2f\left(\edd, \sigma / \xi \right)
- f_{||}\left(\edd, \sigma / \xi \right)/2$.
For small $\edd$, we get $f_{||,\bot}\left(\edd, \sigma / \xi\right)=
A_{||,\bot}(\sigma / \xi)
+B_{||,\bot}(\sigma / \xi)\,\edd+\ldots\,$, again with coefficients $A_{||,\bot}
(\sigma / \xi)$ and $B_{||,\bot} (\sigma / \xi)$, whose absolute values decrease
with increasing disorder correlation length. 
This time, however, only $f_{\bot} \left(\edd, \sigma / \xi \right)$  diverges for
$\edd \uparrow 1$,
whereas $f_{||} \left(\edd, \sigma / \xi \right)$ remains finite in this limit.

Comparing Figs.~\ref{sup} a) and~\ref{sup} b) by taking into account the different
vertical scales, 
we conclude that the superfluid density parallel 
to $\hat{\ve{e}}_z$ is always less depleted
than the superfluid density perpendicular to $\hat{\ve{e}}_z$.  This result can be
qualitatively explained by 
considering $V_{\text{int}}(\ve{k})$, including the dipolar interaction,
as an effective 
contact interaction
strength for a flow with wave vector $\ve{k}$. This quantity is 
smaller than $g$ perpendicular to $\hat{\ve{e}}_z$ and larger than $g$ parallel to
$\hat{\ve{e}}_z$. As the Huang-Meng
depletion $\nhm$ \cite{huang} for a pure contact interaction scales
with the power $-1/2$ with respect to the contact interaction strength, 
we obtain in this picture a smaller depletion parallel to $\hat{\ve{e}}_z$ than
perpendicular 
to $\hat{\ve{e}}_z$.
In addition, we note that $f_{||}\left( 0 , \sigma / \xi \right) > f \left( 0 ,
\sigma / \xi \right)$  \cite{kobayashi},
whereas $f_{||}\left(\edd, \sigma / \xi \right)$ decreases and $f \left(\edd, \sigma
/ \xi \right)$ increases, respectively,
with $\edd$ for fixed  $\sigma / \xi$.
Therefore, a system with a sufficiently large relative interaction strength $\edd$
has the astonishing property that the depletion of the parallel superfluid density
is smaller 
than the condensate depletion.
Figure \ref{sup} a) reveals in the dashed line how the critical value
$\epsilon_{\text{dd,crit}}$
decreases with increasing
disorder correlation length $\sigma$.
The existence of $\epsilon_{\text{dd,crit}}$
represents a counterintuitive result, as particles of the fragmented BECs, which are
supposed to be localized 
in the respective minima of the random potential, seem to
contribute to the parallel superfluid motion of the system. We conclude that, due to
the 
presence of the dipolar interaction, 
the locally condensed particles are only localized for a certain time scale. For
longer time periods, 
our finding suggests that  
an exchange of the localized particles occurs with the nonlocalized 
particles, thus allowing for a 
superfluid density that is larger than the condensate density. 
This supports the finding of Ref.~\cite{graham}, in which such a finite localization
time 
for tiny BECs was calculated within a Hartree-Fock theory of dirty bosons 
for the special case of a pure contact interaction.\\
\begin{figure*}[t]
\includegraphics[scale=0.95]{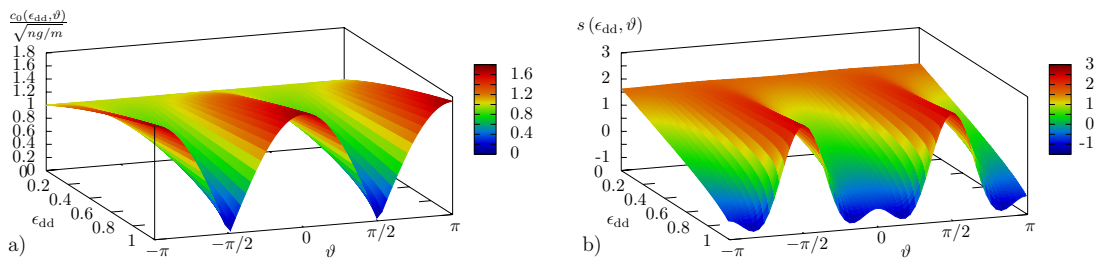}
\caption{(Color online) a) Speed of sound for vanishing disorder and b) correction
term $s(\edd,\vartheta)$ for
nonvanishing disorder in Eq.~(\ref{corr}).}
\label{fig:soundw}
\end{figure*}
Finally, we determine the speed of sound within a hydrodynamic approach
\cite{khalatnikov}.
To this end we assume that, in the long-wavelength limit, the respective transport
quantities
are effectively disorder averaged. Thus, we consider both the Euler equation 
$m \partial  \ve{v}_\text{S} / \partial t
+ \nabla \left(\mu+ m \ve{v}_\text{S}^2 / 2 \right)=0$ and the continuity equation 
$\partial n / \partial t + \nabla\cdot\ve{j}= 0$ with the 
current density $\ve{j} = n_\text{S} \ve{v}_\text{S}+n_\text{N}\ve{v}_\text{N}$,
where $\ve{v}_\text{S}$ and $\ve{v}_\text{N}$ denote  
the superfluid and boost velocity, respectively. As the normal component is pinned,
we must
also assume that it remains stationary. This leads to the condition
$\ve{v}_\text{N}={\bf 0}$, which
is reminiscent of the physically closely related problem of fourth sound in $^4$He
\cite{khalatnikov}.
Considering small space- and time-dependent perturbations around 
the equilibrium values $\ve{v}_\text{S} = \delta \ve{v}_\text{S}(\ve{x},t)$, $n =
n_\text{eq} 
+ \delta n(\ve{x},t)$, $n_\text{S} = n_{\text{S},\text{eq}}
+\delta n_{\text{S}}(\ve{x},t)$,  and $\mu = \mu[n_\text{eq}+\delta n(\ve{x},t)]$ 
yields 
\begin{eqnarray}
\hspace*{-3mm}\frac{\partial^2\delta 
n(\ve{x},t)}{\partial t^2}- \nabla \left[ \frac{n_{\text{S},\text{eq}}}{m}
\left.\frac{\partial \mu}{\partial n}
\right|_\text{eq} \nabla \delta n(\ve{x},t) \right]= 0\,. 
\end{eqnarray}
With this, we obtain in the sound wave regime for the speed of sound in the
direction $\hat{\bf q}$ the general result
\begin{eqnarray}
\hspace*{-5mm}\frac{c}{\sqrt{n V_{\text{int}}(\ve{0})/m}} = 1 + 2 \int \frac{d^3
k}{(2 \pi)^3} 
\frac{R({\bf k})}{\left[ \hbar^2\ve{k}^2 / 2m +2nV_{\text{int}}(\ve{k})\right]^2}
\nonumber \\
\hspace*{-5mm}\times \left\{ 
\frac{\hbar^2\ve{k}^2  V_{\text{int}}(\ve{k}) / 2m
V_{\text{int}}(\ve{0})}{\hbar^2\ve{k}^2 / 2m +2nV_{\text{int}}(\ve{k}) }
- (\hat{\bf q} \, \hat{\bf k})^2   
\right\} +\ldots \,,\,
\end{eqnarray}
where the first term originates from the equation of state and the second term
originates from the superfluid density
(\ref{ns}). Note that our hydrodynamic derivation of the speed of sound reduces for
the special case of a contact
interaction to an expression that has recently been obtained within an independent
Green's function approach
\cite{gaul2}. In the case of a dipolar interaction and an uncorrelated disorder
$R(\ve{x}) = R\delta(\ve{x})$,
we have
\begin{eqnarray}
\frac{c(\edd,\vartheta)}{c_0(\edd,\vartheta)} = 1
+ \frac{\nhm g /m}{2 c_0^2(\edd,\vartheta)}\, s(\edd,\vartheta) +\ldots\,,
\label{corr}
\end{eqnarray}
with $c_0(\edd,\vartheta)= \sqrt{n g (3\edd\cos^2{\vartheta}+1-\edd)/m}$
representing the speed of sound for vanishing disorder
and with a dipolar function $s(\edd,\vartheta)$,
which generalizes the result for the pure contact interaction $s(0,\vartheta) = 5 /
3$ \cite{giorgini}. 
Figure \ref{fig:soundw} shows that due to weak disorder, the speed of sound has a
strong dependence on both the sound direction, characterized by
$\hat{\bf q}\cdot \hat{\bf e}_z = \cos \vartheta$, and the relative interaction
strength $\edd$.  

Finally, we conclude that the delicate interplay of dipolar interaction and weak
disorder yields
characteristic anisotropies for physical observables at $T=0$. The anisotropy of the
speed of sound should be detectable with modern Bragg spectroscopy by measuring
the underlying dynamic structure factor
\cite{ketterle,davidson}. Even more important could be the anisotropy for the
superfluidity density,
as this should affect the collective excitations \cite{falco2} for harmonically
trapped dipolar condensates within a random
environment. It is expected that these interesting effects are more pronounced for
strong dipolar
interactions that arise for highly magnetic atoms, such as dysprosium
\cite{dysprosium}, or polar
heteronuclear molecules, such as $^{40}$K$^{87}$Rb \cite{ospelkaus}. To this end, it
is indispensable to investigate
dirty dipolar BECs in confined geometries, for instance in a harmonic trap
\cite{falco3} or an optical
lattice \cite{muruganandam}.

We thank J. Dietel, R.~Graham, M.~von Hase, P.~Navez, and F. Nogueira
for useful discussions and acknowledge
financial support from the German National Scholarship
and from the German Research Foundation
(DFG) within the Collaborative Research Center SFB/TR12.
\end{document}